\documentclass[12pt]{article}
\usepackage{amssymb}

\input tcilatex

\begin{document}

\title{A boundary-value problem for cold plasma dynamics}
\author{Thomas H. Otway\thanks{%
email: otway@ymail.yu.edu} \\
\\
\textit{Departments of Mathematics and Physics,}\\
\textit{Yeshiva University, 500 W 185th Street,}\\
\textit{New York, New York 10033}}
\date{}
\maketitle

\begin{abstract}
A weak Guderley-Morawetz problem is formulated for a mixed
elliptic-hyperbolic system that arises in models of wave propagation in cold
plasma. \ Weak solutions are shown to exist in a weighted Hilbert space. \
This result extends work by Yamamoto. \textit{MSC2000}: 35M10.
\end{abstract}

\section{Introduction}

A characteristic feature of wave propagation in cold plasma is the
possibility that a hybrid resonance surface, along which the linearized
equation for the electric field changes from elliptic to hyperbolic type,
may be parallel to a flux surface. \ This property can be represented in two
dimensions by setting the hybrid resonance curve tangent to the line $x=0$
at the origin of coordinates. \ The situation is somewhat different from
that found in, for example, linear models of transonic fluid dynamics [eq.
(3), below]. In that case the sonic line is everywhere normal to the line $%
x=0$.

A model for such a resonance curve is the equation 
\[
x=\sigma \left( y\right) , 
\]
where $\sigma \left( y\right) $ is a continuously differentiable function of
its argument satisfying 
\[
\sigma \left( 0\right) =\sigma ^{\prime }\left( 0\right) =0. 
\]
In addition, we assume for simplicity that both $\sigma \left( y\right) \ $%
and $\sigma ^{\prime }\left( y\right) $ exceed zero for $y\ $exceeding zero.

This leads us to consider mixed elliptic-hyperbolic systems having the form 
\begin{equation}
Lu=f,
\end{equation}
where 
\[
u=\left( u_{1}\left( x,y\right) ,u_{2}\left( x,y\right) \right) ,\;f=\left(
f_{1}\left( x,y\right) ,f_{2}\left( x,y\right) \right) , 
\]
\[
\left( x,y\right) \in \Omega \subset \Bbb{R}\times \left( \Bbb{R}/\Bbb{R}%
^{-}\right) , 
\]
\[
\left( Lu\right) _{1}=\left[ x-\sigma \left( y\right) \right]
u_{1x}+Ku_{1}+u_{2y}, 
\]
\[
\left( Lu\right) _{2}=u_{1y}-u_{2x} 
\]
with data 
\begin{equation}
u_{1}\frac{dx}{ds}+u_{2}\frac{dy}{ds}=0
\end{equation}
given on a portion of the boundary of $\Omega $. \ Here $K$ is a constant in 
$\left[ 0,1\right] $ and $ds$ denotes the line element on $\partial \Omega $%
. \ The system is elliptic for $x>\sigma \left( y\right) $ and hyperbolic
for $x<\sigma \left( y\right) $. \ Following [\textbf{MSW}], we emphasize
the analogy to fluid dynamics by calling the curve $x=\sigma \left( y\right) 
$ the \textit{sonic curve}.

In the cold plasma literature, eqs. (1), (2) tend to appear in scalar-valued
special cases. \ In all these cases $\sigma \left( y\right) $ is
proportional to $y^{2},$ but this specific restriction is not imposed by the
physics; concerning the physical model, see [\textbf{W1}], [\textbf{W2}]. \
If $u_{1}=\psi _{x}$, $u_{2}=\psi _{y}$, $\sigma \left( y\right) =y^{2},$
and $f=\left( 0,0\right) $, the system reduces to a scalar equation
introduced in Sec. 3 of [\textbf{MSW}]. \ In the context of this equation,
condition (2) corresponds to imposing constant boundary conditions on the
scalar solution $\psi \left( x,y\right) $. \ A uniqueness theorem was proven
in [\textbf{MSW}] for $K=1/2$, in order to show the existence of a domain on
which the classical Dirichlet problem is ill-posed for the equation. \
Numerical arguments for a complex perturbation were also introduced. \ A
similar equation [$\sigma \left( y\right) \propto y^{2},$ $u_{1}=\psi _{x}$, 
$u_{2}=-\psi _{y}$, $K=1$, $f=\left( 0,0\right) $] appeared earlier in the
physics literature, also in the context of wave propagation in cold plasma [%
\textbf{PF}]. \ In this case certain exact solutions were constructed. \
Finally, system (1), (2) in the case $\sigma \left( y\right) =y^{2}$ was
studied in an interesting unpublished Ph.D. dissertation [\textbf{Y}] on the
existence of weak solutions possessing Dirichlet data on a certain small
domain near the origin of $\Bbb{R}^{2}$.

These equations share with the \textit{Tricomi equation} 
\begin{equation}
y\psi _{xx}+\psi _{yy}=0
\end{equation}
a multiplicity of possible approaches to formulating boundary-value
problems. \ From a mathematical point of view, the Dirichlet problem, in
which data are assigned on the entire boundary, is the ``wrong problem'' to
solve for equations of mixed type, as this problem tends to become
over-determined in the hyperbolic region. \ Well-posed problems for
elliptic-hyperbolic equations generally include a characteristic gap on
which data have not been prescribed. \ In addition to the example of an
over-determined Dirichlet problem for the scalar equation considered in [%
\textbf{MSW}], there are analogous examples for Tricomi-like equations [%
\textbf{M2}].

However, physical applications of eq. (3) to transonic fluid dynamics, and
of scalar forms of (1) to wave propagation in cold plasma, suggest that it
should be possible to prescribe data over the entire boundary. \ This
contradiction suggests that classical solutions will have little application
to such physical problems. \ In terms of weak solutions $u$ to the system
(1), the Dirichlet problem requires solutions to lie in an appropriate
Hilbert space $\widetilde{H_{1}}$ and satisfy 
\[
-\left( u,L^{\ast }\varphi \right) _{L^{2}}=\left( f,\varphi \right) _{L^{2}}
\]
$\forall \varphi =\left( \varphi _{1},\varphi _{2}\right) \in \widetilde{%
H_{2}},$ where $\widetilde{H_{2}}$ is another appropriate Hilbert space and
where $\varphi _{1}=0$ on the entire boundary of the domain [\textit{c.f}. [%
\textbf{M4}], eqs. (6)-(8), for the Tricomi case]. \ This problem may or may
not be well-posed. \ A different approach is to require that the component $%
\varphi _{1}$ of the test function $\varphi $ vanish only on the
noncharacteristic part of the boundary, and that $\varphi $ satisfy
condition (2) on characteristics. \ In the early literature (see, \textit{%
e.g.}, Sec. 4 of [\textbf{M3}]) an elliptic-hyperbolic problem having
Dirichlet data given on the entire boundary is called the \textit{closed,}
or \textit{full} Dirichlet problem to distinguish it from the mathematically
natural case of Dirichlet data given off of the characteristics. \ Following
[\textbf{LP}], we prefer instead to distinguish the problem in which data
are given on the noncharacteristic part of the boundary by calling such a
problem \textit{Guderley-Morawetz}, reserving the term \textit{Dirichlet
problem} for Dirichlet data given on the entire boundary as in [\textbf{M4}]
and [\textbf{P}].

The existence of weak solutions to a Guderley-Morawetz problem is proven for
a Tricomi-like system in [\textbf{M1}]. \ The estimates in Sec. 2.5 of [%
\textbf{Y}] can be extended to imply the existence of weak solutions to a
Guderley-Morawetz problem for eqs. (1), (2) on a relatively large and
general domain. \ This is the content of Theorems 2 and 3 of Sec. 3, below.
\ The arguments in [\textbf{Y}] assume that the weak Dirichlet problem and
the weak Guderley-Morawetz problems are identical. \ (Concerning this
identification, see the Remark following the proof of Proposition 1, below.)
\ It is claimed there, on the basis of Guderley-Morawetz estimates modeled
on [\textbf{M1}], that weak solutions of (1) exist for Dirichlet data
prescribed on the entire boundary. \ We make no such claim for the
generalization of those estimates given here. \ However, the techniques used
to prove the existence of weak solutions to a Guderley-Morawetz problem for
(1), (2) will yield a uniqueness theorem for strong solutions to the
Dirichlet problem for this system, on a more restricted domain, almost ``for
free''\ (see [\textbf{Y}], Theorem 1, Sec. 2.7; also see our Remark, Sec.
3.3). \ Moreover, it is possible to derive the existence of weak solutions
to the Dirichlet problem for eq. (3) by considering a sequence of
Guderley-Morawetz problems in which the characteristic gap is ``marched'' to
a singular point on the sonic curve; see [\textbf{G}], [\textbf{M3}], and [%
\textbf{P}]. \ It is a reasonable conjecture that this method can be
modified to apply to systems such as (1), (2) on an appropriate domain, but
this is not attempted here.

Equations (1) cannot be mapped into a system of the form studied in [\textbf{%
M1}] on any domain that includes the origin. \ On the one hand, relatively
little is known about such elliptic-hyperbolic systems which do not directly
generalize the Tricomi equation. \ On the other hand, the method of proof
adopted here is by now quite standard. \ It is required to find a Hilbert
space $U$, a domain $\Omega ,$ and a multiplier $M$ under which weak
solutions can be shown to exist, without unreasonable restrictions on
generality, by a uniqueness-plus-projection argument using the $abc$ method
and the Riesz Representation Theorem. \ Because this system comes from a
physical model, we additionally hope that our conditions on $U$ and $\Omega $
will be physically reasonable. \ For instance, physical/numerical arguments
for special cases suggest that a singularity should be permitted at the
origin [\textbf{MSW}]. \ This influences the weighting of the Hilbert space $%
U$, as does the existence of particular physical solutions (Sec. 2.2.2).

We note that every nondegenerate conic section is equivalent under the
projective group to the unit circle. \ In that sense, the system (1), (2)
with the choice $\sigma \left( y\right) =y^{2}$ is ``gauge equivalent'' to a
system on the extended projective disc studied in [\textbf{O}] by similar
methods. \ In that system, the elliptic part of the domain has a geometric
interpretation as hyperbolic points in projective space and the hyperbolic
part, as ideal points. \ As it is not clear that projective invariance has
any physical meaning in the context of cold plasma dynamics, this analogy
will not be pursued.

\section{Formulation of the boundary-value problem}

\subsection{Domain}

In proving weak existence for the Guderley-Morawetz problem we assume that
the domain $\Omega ,$ having piecewise continuous boundary, is enclosed by
the arbitrarily large but finite rectangle 
\[
R=\left\{ \left( x,y\right) |-\infty <p\leq x<\ell ,\,0\leq y\leq q<\infty
\right\} , 
\]
where $\ell ,$\ $p,q$ are fixed but arbitrary real constants.\ We assume
that $R$ has been chosen so that the distance along the $x$-axis from $%
\sup_{\Omega }x$ to $\ell $ is an arbitrary but fixed positive number $%
\delta .$ \ Because we are assuming the existence of an elliptic region for
eqs. (1), we take $\ell >0.$

The elliptic region of $\Omega $ consists of the region of the first
quadrant bounded by the sonic curve $x=\sigma \left( y\right) $ and a smooth
curve $C_{1}$ emerging from the origin, along which 
\[
\frac{dy}{dx}\geq 0 
\]
with equality only at the origin, and 
\begin{equation}
a\left( y\right) \frac{dy}{dx}+b\left( y\right) <0
\end{equation}
for specified functions $b(y)\leq 0$ and $a(y)\geq 0$. \ We assume that $%
C_{1}$ intersects the sonic curve at a point $\left( x_{0},\sigma \left(
x_{0}\right) \right) \in \Omega ,$ where $x_{0}>0.$ \ For example, if $a$, $%
b $, and $\sigma $ are defined as in Theorem 3 of Sec. 3, then the family of
curves given by $y=\varepsilon x^{m}$ for $m>1/2$ and $x\geq 0$ satisfies
condition (4) for 
\[
\varepsilon \leq \sqrt{\frac{\ell ^{1-2m}}{Km}} 
\]
whenever $K>0.$ \ [Condition (4) is automatically satisfied for such $a$, $b$%
, and $\sigma $ if $K=0$.] \ If we further specify $m\leq 1/K,$ then we are
guaranteed that $x_{0}\leq \ell -\delta $ provided we choose 
\[
\varepsilon \geq \sqrt{\left( \ell -\delta \right) ^{1-2m}}. 
\]

The hyperbolic region is bounded by a piecewise smooth curve $\Gamma \cup
C_{2},$ where:

$\Gamma $ is a characteristic curve 
\begin{equation}
\frac{dx}{dy}=-\sqrt{\sigma \left( y\right) -x}
\end{equation}
emerging from the sonic curve at $\left( x_{0},\sigma \left( x_{0}\right)
\right) $;

$C_{2}$ is a piecewise continuous curve intersecting the characteristic $%
\Gamma $ at a single point on the left endpoint of $C_{2}$ and intersecting $%
C_{1}$ at the origin on the right endpoint of $C_{2}$. We assume that $%
dy\leq 0$ and $dx\geq 0$ on $C_{2}$. \ We orient the boundary in the
counterclockwise direction.

The relation of this domain to the domain considered in [\textbf{Y}] is
discussed at the end of Section 3.3.

\subsection{Function spaces}

\subsubsection{Weak solutions}

Denote by $U$ the Hilbert space consisting of all pairs of measurable
functions $\left( u_{1},u_{2}\right) $ such that 
\[
\left\| u\right\| _{\ast }=\left[ \int \int_{\Omega }\sigma ^{\prime }\left(
y\right) \left( u_{1}^{2}+u_{2}^{2}\right) dxdy\right] ^{1/2} 
\]
is finite. \ Here 
\[
\left( u,w\right) _{\ast }=\int \int_{\Omega }\sigma ^{\prime }\left(
y\right) \left( u_{1}w_{1}+u_{2}w_{2}\right) dxdy. 
\]
Denote by $W$ the linear space of continuously differentiable functions $%
\left( w_{1},w_{2}\right) $ vanishing at the origin of $\Bbb{R}^{2}$ and
satisfying: 
\begin{equation}
w_{1}dx+w_{2}dy=0
\end{equation}
on the characteristic $\Gamma $, $w_{1}=0$ on $\partial \Omega /\Gamma $,
and 
\[
\left\{ \int \int_{\Omega }\frac{1}{\sigma ^{\prime }\left( y\right) }\left[
\left( L^{\ast }w\right) _{1}^{2}+\left( L^{\ast }w\right) _{2}^{2}\right]
dxdy\right\} ^{1/2}<\infty , 
\]
where 
\[
\left( L^{\ast }w\right) _{1}=\left[ x-\sigma \left( y\right) \right]
w_{1x}+\left( 1-K\right) w_{1}+w_{2y} 
\]
and 
\[
\left( L^{\ast }w\right) _{2}=w_{1y}-w_{2x}. 
\]

We define a \textit{weak solution} to eq. (1) under the boundary condition
(2) to be any $u\in U$ such that $\forall w\in W,$%
\begin{equation}
\left( w,f\right) =-\left( L^{\ast }w,u\right)
\end{equation}
under the $L^{2}$ inner product $\left( \;,\;\right) $.

Denote by $H$ the Hilbert space of measurable functions $\left(
h_{1},h_{2}\right) $ for which the norm 
\[
\left\| h\right\| ^{\ast }=\left[ \int \int_{\Omega }\frac{1}{\sigma
^{\prime }\left( y\right) }\left( h_{1}^{2}+h_{2}^{2}\right) dxdy\right]
^{1/2} 
\]
is finite. \ The inner product on $H$ is given by 
\[
\left( h,g\right) ^{\ast }=\int \int_{\Omega }\frac{1}{\sigma ^{\prime
}\left( y\right) }\left( h_{1}g_{1}+h_{2}g_{2}\right) dxdy. 
\]
The prescribed data $f$ will be assumed to lie in the space $H$.

\subsubsection{Similarity solutions}

Analysis of scalar special cases of system (1) with $\sigma \left( y\right)
=y^{2}$ suggests the presence of a singularity at the point $x=y=0$. \ See,
for example, [\textbf{MSW}], where this is discussed in detail. \ This would
suggest a radial weight for the energy functional. \ Quadratic radial
weights are applied in [\textbf{Y}]. \ However, the space $U$ constructed
here also arises naturally in connection with eqs. (1).

As a simple example, consider similarity solutions for the case $\sigma
\left( y\right) =y^{2}$ having the form $u_{1}=\psi _{x},$ $u_{2}=\psi _{y},$
and 
\[
\psi \left( x,y\right) =x^{\nu }F\left( \frac{y^{2}}{x}\right) , 
\]
where $\nu $ is a parameter and $F$ satisfies the hypergeometric equation 
\[
\left( 1-\mu \right) \left[ \nu \left( \nu -1\right) F\left( \mu \right)
-2\left( \nu -1\right) \mu F^{\prime }\left( \mu \right) +\mu ^{2}F^{\prime
\prime }\left( \mu \right) \right] 
\]
\[
+\left[ 2F^{\prime }\left( \mu \right) +4\mu F^{\prime \prime }\left( \mu
\right) \right] =0 
\]
for 
\[
\mu =\frac{y^{2}}{x}. 
\]
Properties of such solutions for complex values of $\nu $ are studied in [%
\textbf{MSW}]. \ We consider here the case of real-valued $\nu ,$ as in [%
\textbf{PF}] and [\textbf{W1}]. \ It has been observed [\textbf{W1}] that if 
$\left| x\right| $ is sufficiently small, then $F\sim \mu ^{\nu }$ or $F\sim
\mu ^{\nu -1}.$ \ Taking $F\sim \mu ^{\nu }$ for $\nu =1/4,$ we find that 
\[
\psi \left( x,y\right) =x^{1/4}F\left( \frac{y^{2}}{x}\right) \sim y^{1/2}. 
\]
If $u$ lies in the function space $U$, then $\psi $ has weighted Dirichlet
norm 
\[
E_{U}\left( \psi \right) =\left\| u\right\| _{\ast }^{2}=2\int_{\Omega
}y\left( \psi _{x}^{2}+\psi _{y}^{2}\right) dxdy\sim vol(\Omega )/2. 
\]
In fact, $E_{U}\left( \psi \right) $ is finite on $\Omega $ $\forall \nu
\geq 1/4.$ \ If we include solutions of the form $F\sim \mu ^{\nu -1},$ then 
$E_{U}\left( \psi \right) $ would be finite on $\Omega $ $\forall \nu \geq
5/4.$

\subsection{The weak problem is well-posed}

\begin{proposition}
Any twice continuously differentiable weak solution of the Guderley-Morawetz
problem for (1), (2) on $\Omega ,$ as defined by eq. (7), is a classical
solution.
\end{proposition}

\textit{Proof.} \ We refer to the domain as $\Omega ,$ but the argument also
holds without alteration on much more general domains. \ For $u\in U$ and $%
w\in W,$ integration by parts yields 
\[
\left( u,L^{\ast }w\right) =\int \int_{\Omega }u_{1}\left\{ \left[ x-\sigma
\left( y\right) \right] w_{1x}+\left( 1-K\right) w_{1}+w_{2y}\right\} dxdy 
\]
\[
+\int \int_{\Omega }u_{2}\left( w_{1y}-w_{2x}\right) dxdy= 
\]
\[
-\int \int_{\Omega }\left\{ \left[ x-\sigma \left( y\right) \right]
u_{1x}+Ku_{1}+u_{2y}\right\} w_{1}dxdy 
\]
\[
-\int \int_{\Omega }\left( u_{1y}-u_{2x}\right) w_{2}dxdy 
\]
\[
-\int_{\partial \Omega }\left( w_{1}u_{2}+w_{2}u_{1}\right) dx+ 
\]
\begin{equation}
\int_{\partial \Omega }\left\{ \left[ x-\sigma \left( y\right) \right]
w_{1}u_{1}-w_{2}u_{2}\right\} dy.
\end{equation}
On $\partial \Omega /\Gamma $, $w_{1}=0$, implying that 
\begin{equation}
\left( u,L^{\ast }w\right) _{|\partial \Omega /\Gamma }=-\int_{\partial
\Omega /\Gamma }w_{2}\left( u_{1}dx+u_{2}dy\right) .
\end{equation}
Equations (5) and (6) hold on $\Gamma ,$ implying that 
\[
\left( u,Lw\right) _{|\Gamma }=\int_{\Gamma }-\left(
w_{1}u_{2}+w_{2}u_{1}\right) dx+\left\{ \left[ x-\sigma \left( y\right) %
\right] w_{1}u_{1}-w_{2}u_{2}\right\} dy 
\]
\[
=\int_{\Gamma }u_{1}\left\{ \left[ x-\sigma \left( y\right) \right]
w_{1}-w_{2}\frac{dx}{dy}\right\} dy-u_{2}\left( w_{1}dx+w_{2}dy\right) 
\]
\begin{equation}
=\int_{\Gamma }\left[ x-\sigma \left( y\right) +\left( \frac{dx}{dy}\right)
^{2}\right] w_{1}u_{1}dy=0.
\end{equation}
Substituting eqs. (9) and (10) into (8) and using (7), we obtain 
\[
-\left( w,f\right) =\left( u,L^{\ast }w\right) =-\left( Lu,w\right)
-\int_{\partial \Omega /\Gamma }w_{2}\left( u_{1}dx+u_{2}dy\right) . 
\]
Because this identity must hold for every $w\in W,$ we conclude that the
quantity $u_{1}dx+u_{2}dy$ must equal zero almost everywhere on $\partial
\Omega /\Gamma $. \ Applying the hypothesis that $u$ is twice continuously
differentiable, we complete the proof of Proposition 1.

\bigskip

\textbf{Remark}. \ The value of the 1-form $u_{1}dx+u_{2}dy$ on the
characteristic $\Gamma $ is left undetermined by a definition of weak
solution based on (7), so this argument will not establish the
well-posedness of the weak Dirichlet problem for (1), (2) on $\Omega $
(unless we change the boundary conditions on $w$ to $w_{1}=0$ on $\partial
\Omega $). \ However, classical solutions $u$ of either the Dirichlet
problem or the Guderley-Morawetz problem satisfy (7). \ This ambiguity seems
to be the basis for the attempt in [\textbf{Y}] to identify the weak forms
of the two problems; \textit{c.f.} Sec. 2.3 of [\textbf{Y}].

\section{Results}

\begin{theorem}
Let $K\in \left[ 0,1/2\right] $. \ Let the functions $a(y)$ and $b(y)$ in
condition (4) be given by 
\[
a(y)=K\left[ y+\ell ^{-1}\int_{0}^{y}\sigma \left( t\right) dt\right] 
\]
and 
\[
b(y)=-\left[ 1+\frac{\sigma \left( y\right) }{\ell }\right] . 
\]
For every $f\in H,$ there exists on $\Omega $ a weak solution to system (1)
with the boundary condition (2) given on $\partial \Omega /\Gamma .$\ 
\end{theorem}

\begin{theorem}
The conclusion of Theorem 2 extends to the case $K\in \left[ 0,1\right] $ if
we replace the definitions of $a(y)$ and $b(y)$ by 
\[
a(y)=Ky 
\]
and 
\[
b(y)=-\left[ 1+\frac{\sigma \left( y\right) }{2\ell }\right] , 
\]
and specify $\sigma \left( y\right) =y^{2}.$
\end{theorem}

The proofs of Theorems 2 and 3 modify the argument in [\textbf{M1}]. \ In
addition, we adapt a number of choices made in [\textbf{Y}], which is also
based on [\textbf{M1}]. \ The results follow from an \textit{a priori}
estimate.

\begin{lemma}
Under the hypotheses of either Theorem 2 or Theorem 3, $\exists
k>0\backepsilon \forall w\in W$ we have 
\[
k\left\| w\right\| _{\ast }\leq \left\| L^{\ast }w\right\| ^{\ast }. 
\]
\end{lemma}

\subsection{Proof of Lemma 4}

We prove the lemma by the \textit{abc} method. \ Let 
\[
M=\left[ 
\begin{array}{cc}
a & b \\ 
c & d
\end{array}
\right] , 
\]
where $a$ and $b$ are given by the hypotheses of the theorems; $c$ and $d$
will be chosen. \ Then 
\[
I=\left( L^{\ast }w,Mw\right) = 
\]
\[
\int \int_{\Omega }\left\{ \left[ x-\sigma \left( y\right) \right]
w_{1x}+\left( 1-K\right) w_{1}+w_{2y}\right\} \left( aw_{1}+bw_{2}\right)
dxdy 
\]
\[
+\int \int_{\Omega }\left( w_{1y}-w_{2x}\right) \left( cw_{1}+dw_{2}\right)
dxdy. 
\]
Notice that $a$ and $b$ are defined so that $a_{x}=b_{x}=0$. \ We find as in
[\textbf{Y}], Sec. 2.4, that 
\[
a\left[ x-\sigma \left( y\right) \right] w_{1}w_{1x}=\frac{1}{2}\left(
\left\{ a\left[ x-\sigma \left( y\right) \right] w_{1}^{2}\right\}
_{x}-aw_{1}^{2}\right) ; 
\]
\[
bw_{2}\left[ x-\sigma \left( y\right) \right] w_{1x}= 
\]
\[
\left\{ b\left[ x-\sigma \left( y\right) \right] w_{1}w_{2}\right\}
_{x}-bw_{1}w_{2}-b\left[ x-\sigma \left( y\right) \right] w_{1}w_{2x}; 
\]
\[
aw_{1}w_{2y}=\left( aw_{1}w_{2}\right) _{y}-\frac{1}{2}\left(
aw_{2}^{2}\right) _{x}-a_{y}w_{1}w_{2}+aw_{2}w_{2x}-aw_{1y}w_{2}; 
\]
\[
bw_{2}w_{2y}=\frac{1}{2}\left[ \left( bw_{2}^{2}\right) _{y}-b_{y}w_{2}^{2}%
\right] ; 
\]
\[
cw_{1y}w_{1}=\frac{1}{2}\left[ \left( cw_{1}^{2}\right) _{y}-c_{y}w_{1}^{2}%
\right] . 
\]

Choose 
\[
d=a 
\]
and $c=-\left[ x-\sigma \left( y\right) \right] b$. \ Taking into account
cancellations, we can write $I=I_{1}+I_{2}$, where $I_{2}$ is a line
integral and 
\[
I_{1}=\int \int_{\Omega }\left( \alpha w_{1}^{2}+2\beta w_{1}w_{2}+\gamma
w_{2}^{2}\right) dxdy 
\]
for 
\[
\alpha =\frac{1}{2}\left\{ b_{y}\left[ x-\sigma \left( y\right) \right]
-b\left( y\right) \sigma ^{\prime }\left( y\right) \right\} +\left( \frac{1}{%
2}-K\right) a\left( y\right) , 
\]
\[
\beta =-\frac{1}{2}\left[ a_{y}+Kb(y)\right] , 
\]
and 
\[
\gamma =-\frac{1}{2}b_{y}. 
\]

\textit{Case 1:} \ Under the hypothesis on $K$ in Theorem 2, the coefficient
of $a(y)$ in $\alpha $ is nonnegative, and we can write 
\[
\alpha =\frac{\sigma ^{\prime }\left( y\right) }{2\ell }\left[ 2\sigma
\left( y\right) +\ell -x\right] 
\]
\[
+K\left( \frac{1}{2}-K\right) \left[ y+\ell ^{-1}\int_{0}^{y}\sigma \left(
t\right) dt\right] \geq 
\]
\[
\frac{\sigma ^{\prime }\left( y\right) }{2\ell }\left[ 2\sigma \left(
y\right) +\ell -x\right] \geq \frac{\delta \sigma ^{\prime }\left( y\right) 
}{2\ell }, 
\]
\[
\beta =0, 
\]
and 
\[
\gamma =\frac{\sigma ^{\prime }\left( y\right) }{2\ell }. 
\]
Thus in this case we have 
\[
I_{1}\geq \int \int_{\Omega }\left( \alpha w_{1}^{2}+\gamma w_{2}^{2}\right)
dxdy\geq \frac{\chi }{2\ell }\int \int_{\Omega }\sigma ^{\prime }\left(
y\right) \left( w_{1}^{2}+w_{2}^{2}\right) dxdy, 
\]
where $\chi =\min \left\{ \delta ,1\right\} .$

\textit{Case 2:} \ Under the hypotheses of Theorem 3, we have 
\[
\alpha =\frac{y}{2\ell }\left( 2y^{2}+2\ell -x\right) +\left( \frac{1}{2}%
-K\right) Ky\geq 
\]
\[
\frac{y}{2\ell }\left( 2y^{2}+2\ell -x\right) -\frac{y}{2}= 
\]
\[
\frac{y}{2\ell }\left( 2y^{2}+\ell -x\right) \geq \frac{\delta y}{2\ell }, 
\]
\[
\beta =\frac{Ky^{2}}{4\ell }, 
\]
and 
\[
\gamma =\frac{y}{2\ell }. 
\]
Notice that 
\[
\alpha \gamma -\beta ^{2}\geq \left( \frac{y}{2\ell }\right) ^{2}\left(
2y^{2}+\ell -x\right) -\left( \frac{Ky^{2}}{4\ell }\right) ^{2} 
\]
\begin{equation}
\geq \left( \frac{y}{2\ell }\right) ^{2}\left[ \ell -x+\frac{7}{4}y^{2}%
\right] \geq \delta \left( \frac{y}{2\ell }\right) ^{2}.
\end{equation}
Cauchy's inequality implies that 
\begin{equation}
2\beta w_{1}w_{2}\geq -2\left| \beta \right| \left| w_{1}\right| \left|
w_{2}\right| >-2\sqrt{\alpha }\left| w_{1}\right| \sqrt{\gamma }\left|
w_{2}\right| \geq -\alpha w_{1}^{2}-\gamma w_{2}^{2}
\end{equation}
in $\Omega /\left\{ y=0\right\} .$ \ This already implies that the $W$-norm
of $w$ is positive inside the upper half-plane. \ It remains, however, to
derive an explicit lower bound on the coefficient of $y(w_{1}^{2}+w_{2}^{2})$%
.

We claim that there is a constant $\varepsilon \in \left( 0,1\right) ,$
depending only on $R,$ for which 
\begin{equation}
0\leq \alpha \gamma -\delta \left( \frac{y}{2\ell }\right) ^{2}\leq
\varepsilon \alpha \gamma .
\end{equation}
To establish this claim, note that the left-hand inequality in (13) is
obvious from (11), and the right-hand inequality will be satisfied provided 
\[
\alpha \gamma \left( 1-\varepsilon \right) \leq \delta \left( \frac{y}{2\ell 
}\right) ^{2}. 
\]
Assuming without loss of generality that $y$ exceeds zero (the inequality is
true trivially otherwise), our criterion becomes 
\[
2\left\{ y^{2}+\ell \left[ 1+\left( \frac{1}{2}-K\right) K\right] \right\}
-x\leq \frac{\delta }{1-\varepsilon }. 
\]
Replace the quantity on the left by its largest possible value, given that $%
\left( 1/2-K\right) K\leq 1/16.$ \ Our requirement becomes that $\varepsilon 
$ be chosen sufficiently close to 1 so that 
\[
2q^{2}+\frac{17\ell }{8}-p\leq \frac{\delta }{1-\varepsilon }, 
\]
or 
\[
1-\frac{\delta }{2q^{2}+\frac{17\ell }{8}-p}\leq \varepsilon . 
\]
The quantity on the left exceeds zero, as $\ell -p$ exceeds $\delta .$

Now (11) and (13) imply that inequality (12) can be improved to read 
\[
2\beta w_{1}w_{2}\geq -2\left| \beta \right| \left| w_{1}\right| \left|
w_{2}\right| \geq -2\sqrt{\alpha \gamma -\delta \left( \frac{y}{2\ell }%
\right) ^{2}}\left| w_{1}\right| \left| w_{2}\right| \geq 
\]
\[
-2\sqrt{\varepsilon \alpha \gamma }\left| w_{1}\right| \left| w_{2}\right|
\geq -\sqrt{\varepsilon }\alpha w_{1}^{2}-\sqrt{\varepsilon }\gamma
w_{2}^{2}. 
\]
Thus in this case 
\[
I_{1}\geq \left( 1-\sqrt{\varepsilon }\right) \int \int_{\Omega }\left(
\alpha w_{1}^{2}+\gamma w_{2}^{2}\right) dxdy\geq 
\]
\[
\frac{\left( 1-\sqrt{\varepsilon }\right) }{2\ell }\int \int_{\Omega
}y\left( \delta w_{1}^{2}+w_{2}^{2}\right) dxdy\geq \frac{\chi \left( 1-%
\sqrt{\varepsilon }\right) }{2\ell }\int \int_{\Omega }y\left(
w_{1}^{2}+w_{2}^{2}\right) dxdy. 
\]

The remainder of the proof is identical for either set of hypotheses. \ The
boundary terms resulting from applying Green's Theorem on $\Omega $ are
given by 
\[
I_{2}=-\int_{\partial \Omega }\left( \frac{b}{2}\left\{ w_{2}^{2}-\left[
x-\sigma \left( y\right) \right] w_{1}^{2}\right\} +aw_{1}w_{2}\right) dx+ 
\]
\[
\int_{\partial \Omega }\left( \frac{a}{2}\left\{ \left[ x-\sigma \left(
y\right) \right] w_{1}^{2}-w_{2}^{2}\right\} +b\left[ x-\sigma \left(
y\right) \right] w_{1}w_{2}\right) dy. 
\]

On $C_{2}$, $w_{1}=0,$ $dx\geq 0$ and $dy\leq 0,$ so the signs of $a$ and $b$
imply that 
\[
I_{2|C_{2}}=-\frac{1}{2}\int_{C_{2}}aw_{2}^{2}dy+bw_{2}^{2}dx\geq 0. 
\]

On the characteristic $\Gamma $ we have $x\leq \sigma \left( y\right) ,$ $%
dx\leq 0,$ and $dy\geq 0.$ \ In addition, eqs. (5) and (6) imply that 
\[
w_{2}^{2}+\left[ x-\sigma \left( y\right) \right] w_{1}^{2}=w_{1}^{2}\left[
\sigma \left( y\right) -x\right] +\left[ x-\sigma \left( y\right) \right]
w_{1}^{2}=0. 
\]
We have 
\[
I_{2|\Gamma }=I_{21}+I_{22}, 
\]
where 
\[
I_{21}=-\int_{\Gamma }aw_{1}w_{2}dx+\frac{1}{2}\int_{\Gamma }a\left\{ \left[
x-\sigma \left( y\right) \right] w_{1}^{2}-w_{2}^{2}\right\} dy 
\]
\[
=\int_{\Gamma }\frac{a}{2}\left\{ w_{2}^{2}+\left[ x-\sigma \left( y\right) %
\right] w_{1}^{2}\right\} dy=0 
\]
and 
\[
I_{22}=-\int_{\Gamma }\frac{b}{2}\left\{ w_{2}^{2}-\left[ x-\sigma \left(
y\right) \right] w_{1}^{2}\right\} dx 
\]
\[
+\int_{\Gamma }b\left[ x-\sigma \left( y\right) \right] w_{1}w_{2}dy= 
\]
\[
-\int_{\Gamma }\frac{b}{2}\left\{ w_{2}^{2}+\left[ x-\sigma \left( y\right) %
\right] w_{1}^{2}\right\} dx=0. 
\]

On $C_{1}$ inequality (4) holds; in addition, $dx\geq 0,$ $dy\geq 0,$ and $%
w_{1}=0.$ \ Writing 
\[
\int_{C_{1}}aw_{2}^{2}dy=\int_{C_{1}}aw_{2}^{2}\frac{dy}{dx}dx, 
\]
we find that 
\[
I_{2|C_{1}}=-\frac{1}{2}\int_{C_{1}}aw_{2}^{2}dy+bw_{2}^{2}dx= 
\]
\[
-\frac{1}{2}\int_{C_{1}}\left[ a\frac{dy}{dx}+b\right] w_{2}^{2}dx. 
\]
Inequality (4) implies that the integral on the right is nonnegative.

The preceding arguments establish the lower bound of the lemma.

In order to obtain the upper bound for the inequality of Lemma 4 we reason
in either Case 1 or Case 2 as in [\textbf{M1}], writing 
\[
I=\lim_{\tau \rightarrow 0}\left( \frac{L^{\ast }w}{\sqrt{\sigma ^{\prime
}\left( y\right) +\tau }},\left( \sqrt{\sigma ^{\prime }\left( y\right)
+\tau }\right) Mw\right) \leq 
\]
\begin{equation}
C\left( M\right) \left\| L^{\ast }w\right\| ^{\ast }\left\| w\right\| _{\ast
}.
\end{equation}
The constant $C(M)$ will be a finite positive number provided the functions $%
a$, $b$, $c$, $d$ are bounded. \ The existence of such a bound follows from
the finite character of the constants $p$, $q$, and $\ell .$

We obtain, under the hypotheses of either theorem, the inequality 
\[
C^{\prime }\left( p,q,\ell ,\varepsilon ,\delta \right) \left\| w\right\|
_{\ast }\leq \left\| L^{\ast }w\right\| ^{\ast } 
\]
for $C^{\prime }>0.$ \ This completes the proof of Lemma 4.

\subsection{Proof of Theorems 2 and 3}

Both theorems follow from Lemma 4 by a standard argument. \ An inequality
similar to (14) implies that $\forall w\in W,$ 
\[
\left| \left( w,f\right) \right| \leq c_{0}\left\| L^{\ast }w\right\| ^{\ast
}\left\| f\right\| ^{\ast } 
\]
for a constant $c_{0}$ depending only on $\Omega .$ \ For fixed $f\in H,$
the functional 
\[
G\left( L^{\ast }w\right) \equiv \left( w,f\right) 
\]
on $L^{\ast }w$ can be extended to a bounded linear functional on $H$. \ The
Riesz Representation Theorem then implies the existence of an element $%
h=\left( h_{1},h_{2}\right) \in H$ for which 
\[
\left( w,f\right) =\left( L^{\ast }w,h\right) ^{\ast }. 
\]
Defining $u=\left( u_{1},u_{2}\right) ,$ where 
\[
u_{1}=-\frac{h_{1}}{\sigma ^{\prime }\left( y\right) } 
\]
and 
\[
u_{2}=-\frac{h_{2}}{\sigma ^{\prime }\left( y\right) }, 
\]
we find that $u\in U$ and 
\[
\left( w,f\right) =\left( L^{\ast }w,h\right) ^{\ast }=-\left( L^{\ast
}w,u\right) 
\]
$\forall w\in W,$ which completes the proof.

\subsection{Remark}

By slightly modifying the proof of Lemma 4 it is possible to prove the
uniqueness of strong solutions to a Dirichlet problem on a more restricted
domain. \ Replace $\Omega $ by a domain $\Omega ^{\prime },$ in which $C_{2}$
is replaced by the piecewise linear curve $\lambda _{1}\cup \lambda _{2},$
where $\lambda _{1}$ is a vertical line segment $x=const.<0$, lying in the
interior of $R,$ bounded above by $\Gamma $ and below by $\lambda _{2};$
such vertical lines correspond to flux surfaces in the cold plasma model; $%
\lambda _{2}$ is the segment of the $x$-axis bounded on the left by the line
segment $\lambda _{1}$ and on the right by the line $x=0$. \ The curves $%
C_{1}$ and $\Gamma $ are identically defined on $\Omega $ and $\Omega
^{\prime }.$ \ Let condition (4) be satisfied for $b(y)$ defined as in
Theorem 2 and for 
\[
a(y)=\left( 1-K\right) \left[ y+\ell ^{-1}\int_{0}^{y}\sigma \left( t\right)
dt\right] . 
\]
Let $K$ lie in the interval $\left[ 1/2,1\right] .$ \ Then for every $f\in H$
there exists at most one strong solution in $U$ to eqs. (1) on $\Omega
^{\prime }$ with the boundary condition (2) given on almost all of $\partial
\Omega ^{\prime }.$ \ This conclusion extends to the case $K\in \left[ 0,1%
\right] $ if $b(y)$ and $\sigma \left( y\right) $ are defined as in Theorem
3 and $a\left( y\right) =\left( 1-K\right) y$.

By a \textit{strong solution} of (1) we mean an element $u\in U$ for which
there exists a sequence $u^{\nu }\in U$ such that 
\[
\lim_{\nu \rightarrow \infty }\left\| u^{\nu }-u\right\| _{\ast }=0 
\]
and 
\[
\lim_{\nu \rightarrow \infty }\left\| Lu^{\nu }-f\right\| ^{\ast }=0. 
\]
This strong solution satisfies the boundary condition (2) on almost all of $%
\partial \Omega ^{\prime }$ if in addition 
\[
\int_{\partial \Omega ^{\prime }}\left( u_{1}^{\nu }dx+u_{2}^{\nu }dy\right)
^{2}\left( ds\right) ^{-1}=0, 
\]
where $ds$ is the line element on $\partial \Omega ^{\prime }.$

Suppose that we impose the following additional restrictions and
modifications on the domain $\Omega ^{\prime }:$

the arbitrarily large finite rectangle $R$ in the upper half-plane is
replaced by a sufficiently small circle $R_{0}$ in the upper half-plane,
tangent to the origin;

the line segment $\lambda _{1}$ is chosen to lie sufficiently close to the $%
y $-axis;

the line segment $\lambda _{2}$ is \ replaced by that segment of $R_{0}$
bounded on the left by $\lambda _{1}$ and on the right by the $y$-axis;

the characteristic curve $\Gamma $ is a curve satisfying 
\[
\frac{dx}{dy}=-\sqrt{y^{2}-x}, 
\]
emerging from the parabola $x=y^{2}$ at a point $\left( \widetilde{\delta },%
\widetilde{\delta }^{2}\right) $ sufficiently close to the origin;

condition (4) is satisfied for $b(y)=-(1+y^{2})$ and $a(y)=(1-K)y$.

Then $\Omega ^{\prime }$ becomes identical to the domain $D$ considered in
Ch. 2 of [\textbf{Y}]. \ A uniqueness theorem for solutions of (1), (2),
with $\sigma \left( y\right) =y^{2},$ lying in a radially weighted Hilbert
space over $D$ is given in Sec. 2.7 of [\textbf{Y}].

\bigskip

\textbf{References}

\bigskip

[\textbf{LP}] D. Lupo and K. R. Payne, A dual variational approach to a
class of nonlocal semilinear Tricomi problems, \textit{Nonlinear Diff. Eqs.
Appl.} \textbf{6}(3) (1999), 247-266.

[\textbf{G}] G. Guderley, \textit{The Theory of Transonic Flow}, Pergamon,
Oxford, 1962.

[\textbf{M1}] C. S. Morawetz, A weak solution for a system of equations of
elliptic-hyperbolic type, \textit{Commun. Pure Appl. Math.} \textbf{11}
(1958), 315-331.

[\textbf{M2}] C. S. Morawetz, Non-existence of transonic flow past a
profile, \textit{Commun. Pure Appl. Math.} \textbf{27} (1964), 357-367.

[\textbf{M3}] C. S. Morawetz, Mixed equations and transonic flow, \textit{%
Rend. Mat.} \textbf{25} (1966), 1-28.

[\textbf{M4}] C. S. Morawetz, The Dirichlet problem for the Tricomi
equation, \textit{Commun. Pure Appl. Math.} \textbf{23} (1970), 587-601.

[\textbf{MSW}] C. S. Morawetz, D. C. Stevens, and H. Weitzner, \ A numerical
experiment on a second-order partial differential equation of mixed type, 
\textit{Commun. Pure Appl. Math.} \textbf{44} (1991), 1091-1106.

[\textbf{O}] T. H. Otway, Hodge equations with change of type, \textit{Ann.
Mat. Pura Appl.} (to appear).

[\textbf{P}] K. R. Payne, Boundary geometry and location of singularities
for solutions to the Dirichlet problem for Tricomi type equations, \textit{%
Houston J. Math.} \textbf{23}(4) (1997), 709-731.

[\textbf{PF}] A. D. Piliya and V. I. Fedorov, Singularities of the field of
an electromagnetic wave in a cold anisotropic plasma with two-dimensional
inhomogeneity, \textit{Sov. Phys. JETP} \textbf{33} (1971), 210-215.

[\textbf{W1}] H. Weitzner, \textit{Wave propagation in the cold plasma model}%
, Courant Inst. Math. Sci. Magneto-Fluid Dynamics Div. Report MF-103,
August, 1984.

[\textbf{W2}] H. Weitzner, Lower hybrid waves in the cold plasma model, 
\textit{Commun. Pure Appl. Math.} \textbf{38} (1985), 919-932.

[\textbf{Y}] Y. Yamamoto, \textit{Existence and uniqueness of a generalized
solution for a system of equations of mixed type}, Ph.D. Dissertation,
Polytechnic University of New York, 1994.

\end{document}